\documentclass{article}%
\usepackage{graphicx}
\usepackage{amsmath}
\usepackage{amsfonts}
\usepackage{amssymb}%
\begin{document}

\title{Quadratic Lie algebras and quasi-exact solvability of the two-photon Rabi Hamiltonian}
\author{S. N. Dolya\\{B. Verkin Institute for Low Temperature Physics and }Engineering \\47, Lenin Prospekt, Kharkov 61164, Ukraine.\\E-mail: dolya@ilt.kharkov.ua}
\maketitle

\begin{abstract}
It is proved that the two-photon Rabi Hamiltonian is quasi exactly solvable on
the basis of the two different quadratic Lie algebras.

\end{abstract}

\section{Introduction}

By definition, a quasi-exactly solvable (QES) Hamiltonian $\hat{H}$ \ has a
finite-dimensional matrix representation within a finite-dimensional subspace
of the whole infinite-dimensional Hilbert space. As a result, a partial
algebraization of the spectrum occurs. Correspondingly, a QES Lie algebra is
the Lie algebra of differential operators $\mathfrak{g}\left(  J_{1}%
,J_{2},...,J_{N}\right)  $ that admits finite-dimensional representation in
the subspace $\Re_{N}$ of smooth functions ( if $\psi\in\Re_{N}$ and $J_{n}%
\in$ $\mathfrak{g}$ then $J_{n}\left(  \psi\right)  \in\Re_{N}$ )
\cite{olver}. Then, the QES Hamiltonian belongs to an universal enveloping QES algebra.

In investigation of QES systems, it was implied up to now that such a system
has only one underlying QES algebra. In the present paper, we would like to
pay attention that a given QES Hamiltonian may have more than one QES algebra.
In doing so, different finite-dimensional representations of such QES algebras
$\mathfrak{g}_{\mathbf{1}}$, $\mathfrak{g}_{\mathbf{2}}$, $\mathfrak{g}%
_{\mathbf{3}}$.... cut out different parts of the spectrum of Hamiltonian
$\hat{H}$ and/or correspond to different pieces in the space of its
parameters. This observation not only deepens our understanding of the
mathematical structure of QES systems but is also important for physical
applications since it allows to explore the properties of the spectrum in more detail.

In the present paper, we consider the two-photon Rabi Hamiltonian (TPRH). We
find explicitly two different QES algebras that ensure its quasi-exact
solvability for certain values of parameters. It turns out that both the
algebras are quadratic in contrast to the case of linear $sl(2,C)$ algebra
typical of one-dimensional systems \cite{Turb} - \cite{Enciso}. ( Quadratic
algebras in the QES context were also discussed in Ref. \cite{Brih8} ).

The fact that TPRH is QES was shown in Ref. \cite{Emary}. The works
\cite{Emary}, \cite{d} and the present one deal with different relationships
between the parameters of the system and describe, in general, different parts
of the spectrum for which different QES Lie algebras are relevant. Such a rich
QES structure of TPRH is a separate interesting property of this system.

\section{Two-photon Rabi Hamiltonian}

Rabi Hamiltonian \cite{Rabi} describes a two-level system (atom) coupled to a
single mode of radiation via dipole interaction. It reads
\begin{equation}
H_{\text{R}}=\frac{\tilde{\omega}_{0}}{2}\sigma_{z}+\tilde{\omega}\cdot
b^{+}b+\tilde{g}\left(  b+b^{+}\right)  \cdot\left(  \sigma_{+}+\sigma
_{-}\right)  \label{r}%
\end{equation}
where $b$ and $b^{+}$ are the Bose annihilation and creation operators
respectively ($[b,b^{+}]=1$) and $\sigma_{z}$, $\sigma_{\pm}=\sigma_{x}\pm
i\cdot\sigma_{y}$ are the Pauli matrices. It is often considered in the
rotating-wave approximation in which the terms proportional to $b\sigma_{-}$
and $b^{+}\sigma_{+}$ are omitted in which case it is sometimes called the
Jaynes-Cummings Hamiltonian \cite{Garrison}, \cite{Larson} (in some works, the
latter term applies to the full Hamiltonian (\ref{r}) \cite{Barnett}).

The simplest generalization of such a Hamiltonian taking into account
non-linear optical processes is TPRH:%
\begin{equation}
H=\frac{\tilde{\omega}_{0}}{2}\sigma_{z}+\tilde{\omega}\cdot b^{+}b+\tilde
{g}\left(  b^{2}+\left(  b^{+}\right)  ^{2}\right)  \cdot\left(  \sigma
_{+}+\sigma_{-}\right)  \label{8q}%
\end{equation}

The Hamiltonian (\ref{8q}) describes the same physical system "two-level atom
plus radiation" but emission and absorption of radiation occurs due to a
two-photon process instead of a single-photon ones in the ordinary Rabi
Hamiltonian (\cite{Rabi}). In the present work, we consider the full form of
Hamiltonian (\ref{8q}) without using the rotating wave approximation.

In what follows we will use the representation of the Pauli matrices%
\[
\sigma_{x}=\left(
\begin{array}
[c]{cc}%
{1} & {0}\\
{0} & -{1}%
\end{array}
\right)  \text{, }\sigma_{y}=\left(
\begin{array}
[c]{cc}%
{0} & {i}\\
{-i} & {0}%
\end{array}
\right)  \text{, }\sigma_{z}=\left(
\begin{array}
[c]{cc}%
{0} & {1}\\
{1} & {0}%
\end{array}
\right)  \text{.}%
\]
Then, the spectral problem $H\psi=\tilde{E}\psi$ for Hamiltonian (\ref{8q})
can be represented in the form%

\begin{equation}
\left(
\begin{array}
[c]{cc}%
{L_{+}} & {\omega_{0}}\\
{\omega_{0}} & {L_{-}}%
\end{array}
\right)  \left(
\begin{array}
[c]{c}%
{\varphi_{1}}\\
{\varphi_{2}}%
\end{array}
\right)  =E\left(
\begin{array}
[c]{c}%
{\varphi_{1}}\\
{\varphi_{2}}%
\end{array}
\right)  \text{,} \label{9q}%
\end{equation}
where $L_{\pm}$ $=$ $b^{+}b\pm{g/2}\cdot\left(  \left(  b^{+}\right)
^{2}+b^{2}\right)  $, $\omega_{0}$ $=$ ${\tilde{\omega}_{0}/\left(
2\cdot\tilde{\omega}\right)  }$, $g$ $=$ ${4\cdot\tilde{g}/\tilde{\omega}}$,
$E$ $=$ ${\tilde{E}/\tilde{\omega}.}$

Eliminating the function $\varphi_{2}$ from the system of equations (\ref{9q})
we arrive at the fourth-order equation for the component $\varphi_{1}$ of the
wave function:%

\begin{gather}
L_{1}\varphi_{1}=0\nonumber\\
L_{1}=\left(  L_{-}L_{+}-E\left(  L_{+}+L_{-}\right)  +E^{2}-\omega_{0}%
^{2}\right)  \label{10q}%
\end{gather}

We find the conditions under which the operator $L_{1}$ (\ref{10q}) is QES
and, hence, so is Hamiltonian (\ref{8q}). We would like to draw attention to
the following subtlety. We are dealing with the two-component wave function.
Usually, while considering matrix QES Hamiltonians \cite{Brih3} - \cite{Zhd}
it was assumed that all components of the wave function belong to the
invariant subspace. Now, this is, generally speaking, not the case. We require
that, say, the component $\varphi_{1}$ obey QES equation. However, the second
component $\varphi_{2}=1/\omega_{0}\cdot\left(  E-L_{+}\right)  \varphi_{1}$
does not, in general, belong to the invariant subspace. (We may interchange
the role of $\varphi_{1}$ and $\varphi_{2}$ and require that $\varphi_{2}$
obey the QES equation.)

Now we will show that the operator $L_{1}$ (\ref{10q}) can be represented in
terms of generators of two QES algebras and, therefore, the operator $L_{1}$
itself possesses the same invariant subspaces as the aforementioned
generators. The explicit description of corresponding generators and their
invariant subspaces is given in Appendix \textbf{A}. The fact that the
generators under discussion are quasi-exactly solvable is by no means obvious
and was established within the approach called in \cite{d} QES-extension. It
was shown there that the differential operators $J_{2}^{\pm}$ (\ref{1q}),
$J_{3}^{\pm}$ (\ref{4q}) possess invariant subspaces $\Re_{N}^{2}$ , $\Re
_{N}^{3}$ which are not polynomials and the generators cannot be obtained as a
polynomial deformation of the linear algebra $sl(2,R)$ \cite{Brih8}. It turned
out that they can be expressed in terms of hypergeometric functions \cite{d}.
I use here these results from the previous work \cite{d} with the notations
borrowed from that paper. It is also interesting that the corresponding
algebras are quadratic in terms of aforementioned operators as it is seen from
the commutation relations (\ref{3q}), (\ref{6q}).

Let us use the coordinate representation
\begin{equation}
b=\frac{\partial_{x}+x}{\sqrt{2}},\;b^{+}=\frac{-\partial_{x}+x}{\sqrt{2}}%
\end{equation}
in which $2\cdot L_{\pm}$ $=\left(  \pm g-1\right)  \cdot\partial_{x}^{2}$
$+\left(  1\pm g\right)  \cdot x^{2}-1$. We want to express $L_{1}$
(\ref{10q}) in terms of operators $J_{2}^{-}$, $J_{2}^{+}$ or $J_{3}^{-}$,
$J_{3}^{+}$. As these operators have the second order in derivatives while
$L_{1}$ is of the fourth order, we are seeking for the suitable quadratic
combinations. Using also the gauge freedom, we rely on the relation
\begin{equation}
e^{\beta\left(  x\right)  }\cdot L_{1}\cdot e^{-\beta\left(  x\right)
}=P\left(  J_{2}^{-},J_{2}^{+}\right)  |_{t=\eta\left(  x\right)  }%
\end{equation}
(similarly for $J_{3}^{-}$, $J_{3}^{+}$) where $\beta(x)$ is a gauge function,
$P\left(  z,y\right)  $ $=c_{zz}\cdot z^{2}$ $+c_{yy}\cdot y^{2}$
$+c_{yz}\cdot yz$ $+c_{zy}\cdot zy$ $+c_{z}\cdot z$ $+c_{y}\cdot y$ $+c_{0}$
$-$ is the second order polynomial, $\eta\left(  x\right)  $ is the function
that defines the change of variable. Making a simple choice $\beta\left(
x\right)  $ $=const\cdot x^{2}$, $\eta\left(  x\right)  $ $=const\cdot x^{2}$
we obtain the system of equations for the coefficients of the polynomial
$P\left(  z,y\right)  $. Below we give several solutions which we managed to obtain.

\textbf{I)} The subspace $\Re_{N}^{2}$:%
\begin{equation}
e^{c\cdot x^{2}}L_{1}e^{-c\cdot x^{2}}=\left\vert 4g^{2}\cdot\left(  J_{2}%
^{-}\right)  ^{2}+4g\cdot S_{2}-2g^{2}\cdot J_{2}^{-}+4J_{2}^{+}-\omega
_{0}^{2}-a_{0}\right\vert _{t=\xi\cdot x^{2}} \label{11q}%
\end{equation}

where $a_{0}={\left(  3+4N\right)  \cdot\left(  4N\left(  g^{2}-1\right)
-3+5g^{2}\right)  /4}$,
\begin{equation}
\alpha=-{3/4}\text{, }s=-{1/2}\text{ ( the parameters of subspace }\Re_{N}%
^{2}\text{ )} \label{12q}%
\end{equation}%
\begin{equation}
c=\frac{1}{2}\sqrt{\frac{1+g}{1-g}}\text{ ( the parameter of gauge
transformation )} \label{13q}%
\end{equation}%
\begin{equation}
\xi=\frac{{g}}{\sqrt{1-g^{2}}}\text{ ( the parameter of change of variable )}
\label{14q}%
\end{equation}

\begin{equation}
E=-\frac{{1}}{2}+2\left(  N+1\right)  \cdot\sqrt{1-g^{2}}\text{ ( the energy
of the Hamiltonian (\ref{9q}) ).} \label{15q}%
\end{equation}

\textbf{II)} The subspace $\Re_{N}^{3}$%
\begin{equation}
e^{c\cdot x^{2}}L_{1}e^{-c\cdot x^{2}}=\left\vert 4g^{2}\cdot\left(  J_{3}%
^{-}\right)  ^{2}-4g\cdot S_{3}-2g^{2}\cdot J_{3}^{-}+4J_{3}^{+}-\omega
_{0}^{2}-a_{0}\right\vert _{t=\xi\cdot x^{2}} \label{16q}%
\end{equation}

where $a_{0}={\left(  1+2N\right)  \cdot\left(  2N\left(  g^{2}-1\right)
-1+3g^{2}\right)  /4}$,
\begin{equation}
\alpha=\frac{{3}}{4}-\frac{{N}}{2}\text{, }s={1/2}\text{ ( the parameters of
subspaces }\Re_{N}^{3}\text{ )} \label{17q}%
\end{equation}%
\begin{equation}
c=\frac{1}{2}\sqrt{\frac{1-g}{1+g}}\text{ ( the parameter of gauge
transformation )} \label{18q}%
\end{equation}%
\begin{equation}
\xi=\frac{-{g}}{\sqrt{1-g^{2}}}\text{ ( the parameter of change of variable )}
\label{19q}%
\end{equation}

\begin{equation}
E=-\frac{{1}}{2}+\left(  N+1\right)  \cdot\sqrt{1-g^{2}}\text{ ( the energy of
the Hamiltonian (\ref{9q}) ).} \label{20q}%
\end{equation}

The expressions (\ref{11q}), (\ref{16q}) demonstrate explicitly that
Hamiltonian (\ref{8q}) is indeed QES for corresponding values of parameters.
Using the asymptotic formulas for the hypergeometric function \cite{AS}, one
can check that for the values of parameters (\ref{13q}), (\ref{14q}),
(\ref{18q}), (\ref{19q}) the functions belong to the space $L_{2}\left(
-\infty,+\infty\right)  $. For example, for the subspace $\Re_{N}^{2}$ for
$x\rightarrow\infty$ we have $\varphi_{1}\sim\exp\left(  -\left(
c-\xi\right)  x^{2}\right)  $ \cite{AS} where $c-\xi>0$, $0<g<1$ (\ref{13q}).

Now, the matrix representation of the operator $L_{1}$ (\ref{10q}) follows
from those of $J_{2}^{\pm}$ (\ref{1q}) and $J_{3}^{\pm}$ (\ref{4q}). Below we
list explicit formulas for small values of $N$ . For convenience, we also
include in the subspaces $\Re_{N}^{2}$ , $\Re_{N}^{3}$, the gauge factor
$e^{-cx^{2}}$ from eqs. (\ref{11q}), (\ref{16q}).

\textbf{I)} The subspace $\Re_{0}^{2}$ ($N=0$):%
\begin{equation}
\Re_{0}^{2}=span\left\{  e^{-cx^{2}}%
\begin{array}
[c]{c}%
_{1}F_{1}\left[
\genfrac{}{}{0pt}{}{-3/4}{-1/2}%
;\xi x^{2}\right]
\end{array}
\text{, }e^{-cx^{2}}%
\begin{array}
[c]{c}%
_{1}F_{1}\left[
\genfrac{}{}{0pt}{}{1/4}{1/2}%
;\xi x^{2}\right]
\end{array}
\right\}  \label{21q}%
\end{equation}
where $c$ and $\xi$ are the parameters (\ref{13q}), (\ref{14q}).
Finite-dimensional representations of the operators $J_{2}^{\pm}$ take the
form:%
\begin{equation}
J_{2}^{+}\rightarrow\frac{1}{2}\left(
\begin{array}
[c]{cc}%
{0} & {0}\\
{1} & {-1}%
\end{array}
\right)  \text{, }J_{2}^{-}\rightarrow\frac{1}{4}\left(
\begin{array}
[c]{cc}%
{-3} & {6}\\
{0} & {1}%
\end{array}
\right)  \label{22q}%
\end{equation}
whence, taking into account eq. (\ref{11q}) we obtain the matrix
representation for the operator $L_{1}$ on the subspace $\Re_{0}^{2}$ :%
\begin{equation}
L_{1}\rightarrow\left(
\begin{array}
[c]{cc}%
{3g+{9/4}-\omega_{0}^{2}} & {-3g\left(  2g+1\right)  }\\
{2\left(  g+1\right)  } & {{1/4}-3g-4g^{2}-\omega_{0}^{2}}%
\end{array}
\right)  \label{23q}%
\end{equation}
The spectral problem (\ref{10q}) possesses a non-zero solution if the
determinant of the matrix (\ref{23q}) is equal to zero. The only real solution
$g\left(  \omega_{0}\right)  $ of the corresponding equation that obeys the
condition $0<g<1$ has the form%
\begin{equation}
g\left(  \omega_{0}\right)  =\frac{\sqrt{\left(  4\omega_{0}^{2}-9\right)
\left(  1-4\omega_{0}^{2}\right)  }}{8\omega_{0}}\text{, }\frac{1}{2}%
<\omega_{0}<\frac{3}{2}\text{.} \label{24q}%
\end{equation}
The components of the wave function corresponding to the solution (\ref{24q})
for the particular values of parameters are listed in Appendix \textbf{B. }

\textbf{II)} The subspace $\Re_{2}^{3}$ ($N=2$):%
\begin{equation}
\Re_{2}^{3}=span\left\{
\begin{array}
[c]{c}%
e^{-cx^{2}}%
\begin{array}
[c]{c}%
_{1}F_{1}\left[
\genfrac{}{}{0pt}{}{-3/4}{1/2}%
;\xi x^{2}\right]
\end{array}
\text{, }e^{-cx^{2}}%
\begin{array}
[c]{c}%
_{1}F_{1}\left[
\genfrac{}{}{0pt}{}{3/4}{1/2}%
;\xi x^{2}\right]
\end{array}
\text{,}\\
\text{ }e^{-cx^{2}}%
\begin{array}
[c]{c}%
_{1}F_{1}\left[
\genfrac{}{}{0pt}{}{7/4}{1/2}%
;\xi x^{2}\right]
\end{array}
\end{array}
\right\}  \label{25q}%
\end{equation}
where $c$ and $\xi$ are the parameters (\ref{18q}), (\ref{19q}). Taking into
account (\ref{5q}), (\ref{16q}) we find the finite-dimensional representations
of operators $J_{3}^{\pm}$:%
\begin{equation}
J_{3}^{+}\rightarrow\frac{1}{4}\left(
\begin{array}
[c]{ccc}%
{-2} & {2} & {0}\\
{1} & {2} & {-3}\\
{0} & {10} & {-10}%
\end{array}
\right)  \text{, }J_{3}^{-}\rightarrow\frac{1}{4}\left(
\begin{array}
[c]{ccc}%
{-1} & {0} & {0}\\
{0} & {3} & {0}\\
{0} & {0} & {7}%
\end{array}
\right)  \label{26q}%
\end{equation}
whence, with (\ref{16q}) taken into account, we also obtain the matrix
representation for the operator $L_{1}$ on the subspace $\Re_{2}^{3}$ :%
\begin{equation}
L_{1}\rightarrow\left(
\begin{array}
[c]{ccc}%
{{17/4}-\omega_{0}^{2}-8g^{2}} & {2\left(  g+1\right)  } & {0}\\
{1-g} & {{33/4}-\omega_{0}^{2}-8g^{2}} & {-3\left(  g+1\right)  }\\
{0} & {10\left(  1-g\right)  } & {-{15/4}-\omega_{0}^{2}}%
\end{array}
\right)  \label{27q}%
\end{equation}
The spectral problem (\ref{10q}) has a non-zero solution if the determinant of
the matrix (\ref{27q}) is equal to zero. We find two solutions $g_{\pm}\left(
\omega_{0}\right)  $ satisfying the condition $0<g_{\pm}\left(  \omega
_{0}\right)  <1$:%
\begin{equation}
g_{\pm}\left(  \omega_{0}\right)  =\frac{1}{8}\sqrt{\frac{\left(  5\mp
2\omega_{0}\right)  \left(  2\omega_{0}\pm3\right)  \left(  1\pm2\omega
_{0}\right)  }{\omega_{0}}}\text{, }\frac{1}{2}<\omega_{0}<2\pm\frac{1}%
{2}\text{.} \label{28q}%
\end{equation}
The components of the wave function corresponding to the solution
$g_{+}\left(  \omega_{0}\right)  $ are listed in Appendix \textbf{B} for
particular values of parameters.

Equations (\ref{11q}), (\ref{16q}) obtained above give evidence that
Hamiltonian (\ref{8q}) is quasi-exactly solvable on the basis on two quadratic
Lie algebras (see Appendix \textbf{A}). Each of them cuts out some set of
eigenvalues from the spectrum of the total Hamiltonian. Thus, different parts
of the spectrum are described by different Lie algebras.

\section{Summary and outlook}

It is shown that the two-photon Rabi Hamiltonian (\ref{8q}) is quasi-exactly
solvable on the basis of two different quadratic Lie algebras. The results of
the present paper demonstrate that different parts of the spectrum correspond
to different Lie algebras. It is especially interesting that the functions
which realize invariant subspaces are expressed in terns of special functions
(cf. also \cite{d}) whereas application of the linear Lie algebra leads to
subspaces spanned on polynomials \cite{Emary}.

The results obtained provoke also further questions which we outline here
briefly. First of all, it concerns the comparison of the present article with
previous results. Both in Ref. \cite{Emary} and in our paper QES solutions are
obtained for the energies
\begin{equation}
E=-\frac{1}{2}+M\cdot\sqrt{1-g^{2}} \label{Emary}%
\end{equation}
where $M$ is a half-integer in \cite{Emary} and an integer in our paper. In
\cite{Emary}, these energies were claimed to correspond to level-crossing of
the TPRH but not to all of them. It was conjectured there that integer $M$
might correspond to the remaining crossings. Now, in view of the present
results, it would be especially interesting to elucidate the relationship
between level-crossings and different types of quasi-exact solvability.

In the previous paper \cite{d} we used another representation - the
Fock-Bargmann one instead of the coordinate one in the present work.
Correspondingly, the operators $L_{1}$ and its analogue $\mathbf{L}$ from
\cite{d} do not coincide, so that the conditions of their quasi-exact
solvability impose different restrictions on the parameters of Hamiltonian. It
turned out that in the first case it reduces to $g^{\ast}=\frac{2}{\sqrt{6}}%
$\cite{d} independently of $N$ whereas in our case $g$ is the function of
$\omega_{0}$, the concrete form of this function being different for different
$N$ (see eq. (\ref{24q}) for $N=0$ and (\ref{28q}) for $N=2$).

It is of separate interest the issue of classification of states. In
particular, it concerns the question as to which multiplet the ground state
belongs. If the corresponding relevant QES algebra is known, the problem of
minimization of the functional $\left(  \Psi,\hat{H}\Psi\right)  $ reduces
from the whole Hilbert space to its finite-dimensional part that simplifies
greatly the problem and can be useful for applications.

\section{Acknowledgments}

The author thanks O. B Zaslavskii for useful discussions.

\section{Appendix A.}

Here we give explicit formulas for two QES quadratic Lie algebras used in the
text \cite{d}.

\textbf{I)} Finite dimensional function subspace $\mathcal{R}_{N}%
^{2}=span\{f_{0}^{+},$ $...,$ $f_{N}^{+},$ $f_{0}^{-},$ $...,$ $f_{N}^{-}\}$
$(N=0,1,2,....),$ $\dim\left(  \mathcal{R}_{N}^{2}\right)  =2\left(
N+1\right)  $, formed by functions $f_{n}^{+}=t^{n}\cdot%
\begin{array}
[c]{c}%
_{1}F_{1}\left[
\genfrac{}{}{0pt}{}{\alpha}{s}%
;t\right]
\end{array}
$, $f_{n}^{-}=t^{n}\cdot%
\begin{array}
[c]{c}%
_{1}F_{1}\left[
\genfrac{}{}{0pt}{}{\alpha+1}{s+1}%
;t\right]
\end{array}
$ $(n=0,1,..,N-1,N)$ \cite{AS}, is invariant for the operators $J_{2}^{\pm}$:%
\begin{equation}%
\genfrac{}{}{0pt}{}{J_{2}^{-}=t\frac{d^{2}}{dt^{2}}+\left(  1+s-t\right)
\frac{d}{dt}}{\text{ }J_{2}^{+}=t^{2}\frac{d^{2}}{dt^{2}}+\left(
s-2N-t\right)  \cdot t\frac{d}{dt}+t\left(  N-\alpha\right)  }
\label{1q}%
\end{equation}
The operators $J_{2}^{\pm}$ act on the functions $f_{n}^{\pm}(x)\in
\mathcal{R}_{N}^{2}$ as follows:
\begin{align}
J_{2}^{+}\binom{f_{n}^{+}}{f_{n}^{-}}  &  =\binom{n\left(  A_{n}-1+s\right)
\cdot f_{n}^{+}-B_{n}\cdot f_{n+1}^{+}+2\alpha B_{n}/s\cdot f_{n+1}^{-}%
}{\left(  s-n\right)  \left(  1-A_{n}\right)  \cdot f_{n}^{-}+B_{n}\cdot
f_{n+1}^{-}+s\left(  2B_{n}-1\right)  \cdot f_{n}^{+}}\label{2q}\\
J_{2}^{-}\binom{f_{n}^{+}}{f_{n}^{-}}  &  =\binom{\left(  \alpha-n\right)
f_{n}^{+}+n\left(  n+s\right)  \cdot f_{n-1}^{+}+\alpha\left(  1+2n\right)
/s\cdot f_{n}^{-}}{\left(  \alpha+n+1\right)  f_{n}^{-}+n\left(  n-s\right)
\cdot f_{n-1}^{-}+2ns\cdot f_{n-1}^{+}}\nonumber
\end{align}
where $A_{n}=n-2N$, $B_{n}=n-N$. The commutation relations of the operators
$J_{2}^{\pm}$ are:%
\begin{gather}
{\left[  J_{2}^{+},J_{2}^{-}\right]  }{=S_{2}}\nonumber\\
{\left[  J_{2}^{+},S_{2}\right]  }{=4\cdot J_{2}^{+}J_{2}^{-}-2\cdot
S_{2}-c_{5}^{+}J_{2}^{+}-c_{6}^{+}J_{2}^{-}-c_{7}^{+}}\label{3q}\\
{\left[  J_{2}^{-},S_{2}\right]  }{=-2\cdot\left(  J_{2}^{-}\right)
^{2}-J_{2}^{+}+c_{5}^{+}J_{2}^{-}-\left(  \alpha-N\right)  \left(  s+1\right)
}\nonumber
\end{gather}
where $c_{5}^{+}$ $=$ $2\left(  1+\alpha\right)  +s$, $c_{6}^{+}$ $=$ $\left(
2N-s\right)  \left(  s-2-2N\right)  $, $c_{7}^{+}$ $=$ $\left(  N-\alpha
\right)  \left(  2+2N-s\right)  \cdot$ $\left(  s+1\right)  $.

\textbf{II}) Finite dimensional function subspace $\mathcal{R}_{N}%
^{3}=span\{f_{0},$ $f_{1},$ $...,$ $f_{N}\}$ $(N=0,1,2,....),$ $\dim\left(
\mathcal{R}_{N}^{3}\right)  =N+1$, formed by functions $f_{n}=%
\begin{array}
[c]{c}%
_{1}F_{1}\left[
\genfrac{}{}{0pt}{}{\alpha+n}{s}%
;t\right]
\end{array}
$ $(n=0,1,..,N-1,N)$ \cite{AS}, is invariant for the operators $J_{3}^{\pm}$:
\begin{equation}%
\genfrac{}{}{0pt}{}{J_{3}^{-}=t\frac{d^{2}}{dt^{2}}+\left(  s-t\right)
\frac{d}{dt}}{\text{ }J_{3}^{+}=t^{2}\frac{d^{2}}{dt^{2}}+\left(
s-N-t\right)  \cdot t\frac{d}{dt}-\alpha t}%
\label{4q}%
\end{equation}
The operators $J_{3}^{\pm}$ act on the functions $f_{n}(x)\in\mathcal{R}%
_{N}^{3}$ as follows:
\begin{align}
J_{3}^{-}\left(  f_{n}\right)   &  =\left(  n+\alpha\right)  \cdot
f_{n}\label{5q}\\
J_{3}^{+}\left(  f_{n}\right)   &  =\left(  s\cdot n+C_{n}\right)  \cdot
f_{n}+\alpha_{n}\cdot B_{n}\cdot f_{n+1}+n\left(  \alpha_{n}-s\right)  \cdot
f_{n-1}\nonumber
\end{align}
where $\alpha_{n}=n+\alpha$, $C_{n}=\alpha_{n}\cdot\left(  N-2n\right)  $,
$B_{n}=n-N$ . The commutation relations of the operators $J_{3}^{\pm}$ are:%
\begin{gather}
{\left[  J_{3}^{+},J_{3}^{-}\right]  =S_{3}}\nonumber\\
{\left[  J_{3}^{+},S_{3}\right]  =4\cdot J_{3}^{+}\cdot J_{3}^{-}-2\cdot
S_{3}-c_{5}^{+}\cdot J_{2}^{+}+c_{6}^{+}\cdot J_{2}^{-}+c_{7}^{+}}\label{6q}\\
{\left[  J_{3}^{-},S_{3}\right]  =-2\cdot\left(  J_{3}^{-}\right)  ^{2}%
-J_{3}^{+}+c_{5}^{+}\cdot J_{3}^{-}-s\cdot\alpha}\nonumber
\end{gather}
where $c_{5}^{+}$ $=$ $s+2\alpha+N$ , $c_{6}^{+}$ $=$ $\left(  N+2-s\right)
\cdot$ $\ \left(  N-s\right)  $ , $c_{7}^{+}$ $=$ $s\alpha\cdot$ $\ \left(
N+2-s\right)  $.

The operators $J_{2}^{\pm}\left(  \alpha,s,N\right)  $ (\ref{1q}) can be
transformed into the operators $J_{3}^{\pm}$ $\left(  \alpha^{\prime
},s^{\prime},N^{\prime}\right)  $ (\ref{4q}) by means of substitution of the
parameters $s\rightarrow s^{\prime}-1$ , $\alpha\rightarrow\alpha^{\prime
}+{\left(  N^{\prime}-1\right)  /2}$, $N\rightarrow{\left(  N^{\prime
}-1\right)  /2}$%
\begin{equation}
J_{2}^{\pm}\left(  \alpha^{\prime}+{\left(  N^{\prime}-1\right)  /2}%
,s^{\prime}-1,{\left(  N^{\prime}-1\right)  /2}\right)  =J_{3}^{\pm}\left(
\alpha^{\prime},s^{\prime},N^{\prime}\right)  \text{.} \label{7q}%
\end{equation}
This substitution has meaning for odd $N^{\prime}$ only. It is also worth
noting that the subspace $\Re_{N}^{3}$ is the subset of the subspace $\Re
_{N}^{2}$ ( $\Re_{N}^{3}\subset\Re_{N}^{2}$ ). This claim follows from the
properties of the hypergeometric functions \cite{AS}.

\section{Appendix B.}

\textbf{I)} The subspace $\Re_{0}^{2}$ ($N=0$):

According to eq. (\ref{10q}), we need the eigenvector corresponding to the
zero eigenvalue of the transposed matrix (\ref{23q}) where we should take into
account (\ref{24q}). Then, we obtain for the wave function:%
\begin{equation}
\Psi=\left(
\begin{array}
[c]{c}%
{\varphi_{1}}\\
{\varphi_{2}}%
\end{array}
\right)  =\left(
\begin{array}
[c]{c}%
{p_{11}\cdot h_{1}\left(  x\right)  +p_{12}\cdot h_{2}\left(  x\right)  }\\
{p_{21}\cdot h_{1}\left(  x\right)  +p_{22}\cdot h_{2}\left(  x\right)  }%
\end{array}
\right)  \label{A1}%
\end{equation}
where $h_{1}\left(  x\right)  =e^{-cx^{2}}%
\begin{array}
[c]{c}%
_{1}F_{1}\left[
\genfrac{}{}{0pt}{}{-3/4}{-1/2}%
;\xi x^{2}\right]
\end{array}
$, $h_{2}\left(  x\right)  =e^{-cx^{2}}%
\begin{array}
[c]{c}%
_{1}F_{1}\left[
\genfrac{}{}{0pt}{}{1/4}{1/2}%
;\xi x^{2}\right]
\end{array}
$, $c=\frac{\sqrt{15}+1}{2\left(  \sqrt{15}-1\right)  }$, $\xi=\frac{\sqrt
{15}}{7}$, $p_{11}=250+68\sqrt{15}$, $p_{12}=-235-60\sqrt{15}$, $p_{21}%
=70+28\sqrt{15}$, $p_{22}=35-\left(  120+20\sqrt{15}\right)  x^{2}$. The
second component $\varphi_{2}$ was found from relation (\ref{9q}):%
\begin{equation}
\varphi_{2}=\frac{1}{\omega_{0}}\left(  E-L_{+}\right)  \varphi_{1} \label{A2}%
\end{equation}
The solution (\ref{A1}) (not normalized) corresponds to the energy value
$E=5/4$.

\textbf{II)} The subspace $\Re_{2}^{3}$ ($N=2$)

Proceeding along the same lines as in \textbf{I) }and taking into
account\textbf{ }(\ref{27q})\textbf{ }we obtain the wave function
corresponding to the solution $g_{+}\left(  \omega_{0}\right)  $ at
$\omega_{0}=1$%

\begin{equation}
\Psi=\left(
\begin{array}
[c]{c}%
{\varphi_{1}}\\
{\varphi_{2}}%
\end{array}
\right)  =\left(
\begin{array}
[c]{c}%
{p_{11}\cdot h_{1}\left(  x\right)  +p_{12}\cdot h_{2}\left(  x\right)  }\\
{p_{21}\cdot h_{1}\left(  x\right)  +p_{22}\cdot h_{2}\left(  x\right)  }%
\end{array}
\right)  \label{A3}%
\end{equation}
where $h_{1}\left(  x\right)  =e^{-cx^{2}}%
\begin{array}
[c]{c}%
_{1}F_{1}\left[
\genfrac{}{}{0pt}{}{3/4}{3/2}%
;\xi x^{2}\right]
\end{array}
$, $h_{2}\left(  x\right)  =e^{-cx^{2}}%
\begin{array}
[c]{c}%
_{1}F_{1}\left[
\genfrac{}{}{0pt}{}{-1/4}{1/2}%
;\xi x^{2}\right]
\end{array}
$, $c=\frac{\sqrt{8-3\sqrt{5}}}{2\sqrt{8+3\sqrt{5}}}$, $\xi=-3\sqrt{\frac
{5}{19}}$, $p_{11}=3x^{2}\left(  \left(  7110+3184\sqrt{5}\right)  x^{2}%
-\sqrt{19}\left(  371+170\sqrt{5}\right)  \right)  $, $p_{12}=-\frac{1}%
{4}\left(  \sqrt{19}\left(  6368+2844\sqrt{5}\right)  x^{2}-\left(
1235+418\sqrt{5}\right)  \right)  $, $p_{21}=\frac{57}{2}\left(  79+32\sqrt
{5}\right)  x^{2}$, $\ p_{22}\left(  x\right)  =\frac{19}{4}\left(
76x^{2}-\sqrt{19}\left(  70-31\sqrt{5}\right)  \right)  $. The second
component $\varphi_{2}$ was obtained from relation (\ref{A2}). Solution
(\ref{A3}) (not normalized) corresponds to the energy value $E=3\sqrt
{19}/8-1/2$.

\end{document}